\documentclass{PoS}
\usepackage{amsmath}
\newcommand{\bra}{\langle}
\newcommand{\ket}{\rangle}

\newcommand{\quarter}{\frac{1}{4}}

\newcommand{\order}{{\cal O}}
\newcommand{\One}{1\kern-4.5pt1}
\newcommand{\del}[2]{\frac{\partial#1}{\partial#2}}
\newcommand{\dell}[2]{\partial#1/\partial#2}

\newcommand{\eps}{\varepsilon}

\newcommand{\psibar}{\overline{\psi}}

\newcommand{\eq}{\varepsilon_q}

\newcommand{\qq}{\bra qq\ket}
\newcommand{\pbp}{\bra\psibar\psi\ket}

\newlength{\colw}
\newcommand{\braket}[1]{\langle#1\rangle}
\newcommand{\tr}{\operatorname{Tr}}
\renewcommand{\Re}{\operatorname{Re}}
\newcommand{\err}[2]{\mbox{$\stackrel{\scriptstyle +#1}{\scriptstyle -#2}$}}
\newcommand{\xv}{{\mathbf x}}
\providecommand{\dell}[2]{\partial#1/\partial#2}

\newcommand{\cdeconf}{\cite{Hands:2006ve}}

\newcommand{\cphases}{\cite{Cotter:2012mb}}

\title{Phase structure of QC$_2$D at high temperature and density}

\ShortTitle{Phase structure of QC$_2$2D}

\author{Seamus Cotter, \speaker{Jon-Ivar Skullerud}\\
  Department of Mathematical Physics,
        National University of Ireland Maynooth,\\
	Maynooth, County Kildare,
	Ireland\\
        E-mail: \email{jonivar@thphys.nuim.ie}}

\author{Pietro Giudice, Simon Hands\\ Department of Physics, 
  College of Science,
  Swansea University, Singleton Park, Swansea SA2 8PP, Wales, UK}
\author{Seyong Kim\\Department of Physics, Sejong University,
  Gunja-Dong, Gwangjin-Gu, Seoul 143--747, Korea}

\author{Dhagash Mehta\\Department of Physics, Syracuse University,
Syracuse, NY 13244, USA}

\abstract{We study two-color QCD with two flavors of Wilson fermion as a
function of quark chemical potential $\mu$ and temperature $T$.  We find
evidence of a superfluid phase at intermediate $\mu$ and low $T$ where
the quark number density and diquark condensate are both very well
described by a Fermi sphere of nearly-free quarks disrupted by a BCS
condensate.  This gives way to a region of deconfined quark matter at
higher $T$ and $\mu$, with the deconfinement temperature decreasing
only very slowly with increasing chemical potential. We find that
heavy quarkonium bound states persist in the S-wave channels at all
$T$ and $\mu$, with an energy reflecting the phase structure. P-wave
states appear not to survive in the quarkyonic region.}

\FullConference{The 30th International Symposium on Lattice Field Theory\\
		 June 24 - 29,  2012\\
		 Cairns, Australia}

\begin{document}

\section{Introduction}

Our understanding of the phase structure of QCD at high baryon density
and low temperature remains severely hampered by the sign problem.  In
the absence of first-principles methods which have been proven to
circumvent this problem, we can study a related theory, QCD with
colour group SU(2) (QC$_2$D), which does not suffer from the sign
problem.  This may firstly allow us to confront model studies with
lattice results, thereby constraining these models in their
application to real QCD, and secondly reveal generic features of the
phase structure of strongly interacting gauge theories, including the
nature of deconfinement at high density.

Here we present an update of our ongoing investigation of the phase
structure of QC$_2$D as a function of temperature and chemical
potential \cite{Cotter:2012mb,Hands:2012yy}.

\section{Simulation details}

We study two-colour QCD with a conventional Wilson action for the
gauge fields and two flavours of unimproved Wilson fermion.  The
fermion action is augmented by a gauge- and iso-singlet diquark source
term which serves the dual purpose of lifting the low-lying
eigenvalues of the Dirac operator and allowing a controlled study of
diquark condensation.  Further details about the action and the Hybrid
Monte Carlo algorithm used can be found in \cite{Hands:2006ve}.
We have performed simulations at $\beta=1.9, \kappa=0.168$,
corresponding to a lattice spacing $a=0.178$fm, determined from the
string tension, and a pion mass
$am_\pi=0.645$ or $m_\pi\approx710$MeV.  The ratio of the ground state
pseudoscalar to vector masses is $m_\pi/m_\rho=0.80$ \cite{Hands:2010gd}.
Our lattice volumes and the corresponding values for temperatures $T$,
chemical potentials $\mu$ and diquark sources $j$ are given in
table~\ref{tab:params}.
All results shown are extrapolated to $j=0$ using a linear Ansatz
except where otherwise stated.
\begin{table}[thb]
\begin{center}
\begin{tabular}{|rr|rrr|}
\hline
$N_s$ & $N_\tau$ & $T$ (MeV) & $\mu a$ & $ja$ \\ \hline
16 & 24 & 47 & 0.3--0.9 & 0.04 \\
12 & 24 & 47 & 0.25--1.1 & 0.02, 0.03, 0.04 \\
12 & 16 & 70 & 0.3--0.9 & 0.04 \\
16 & 12 & 94 & 0.2--0.9 & 0.02, 0.04 \\
16 & 8 & 141 & 0.1--0.9 & 0.02, 0.04 \\ \hline
\end{tabular}
\caption{Lattice volumes and associated temperatures $T$, chemical
  potentials $\mu$ and diquark sources $j$.}
\label{tab:params}
\end{center}
\end{table}

\section{Order parameters and phase structure}
\label{sec:phases}

\begin{figure}[tb]
\includegraphics*[width=\colw]{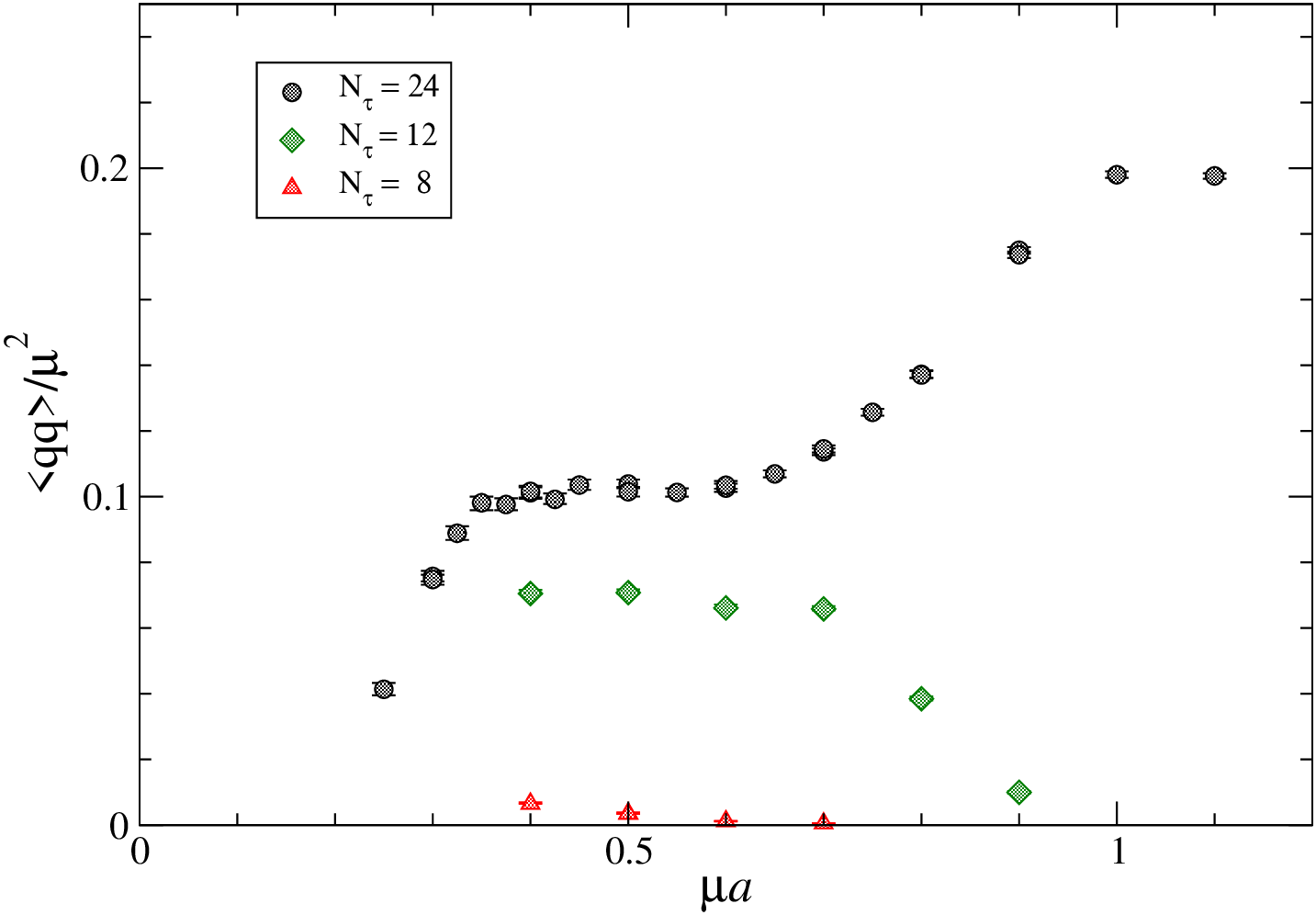}
\includegraphics*[width=\colw]{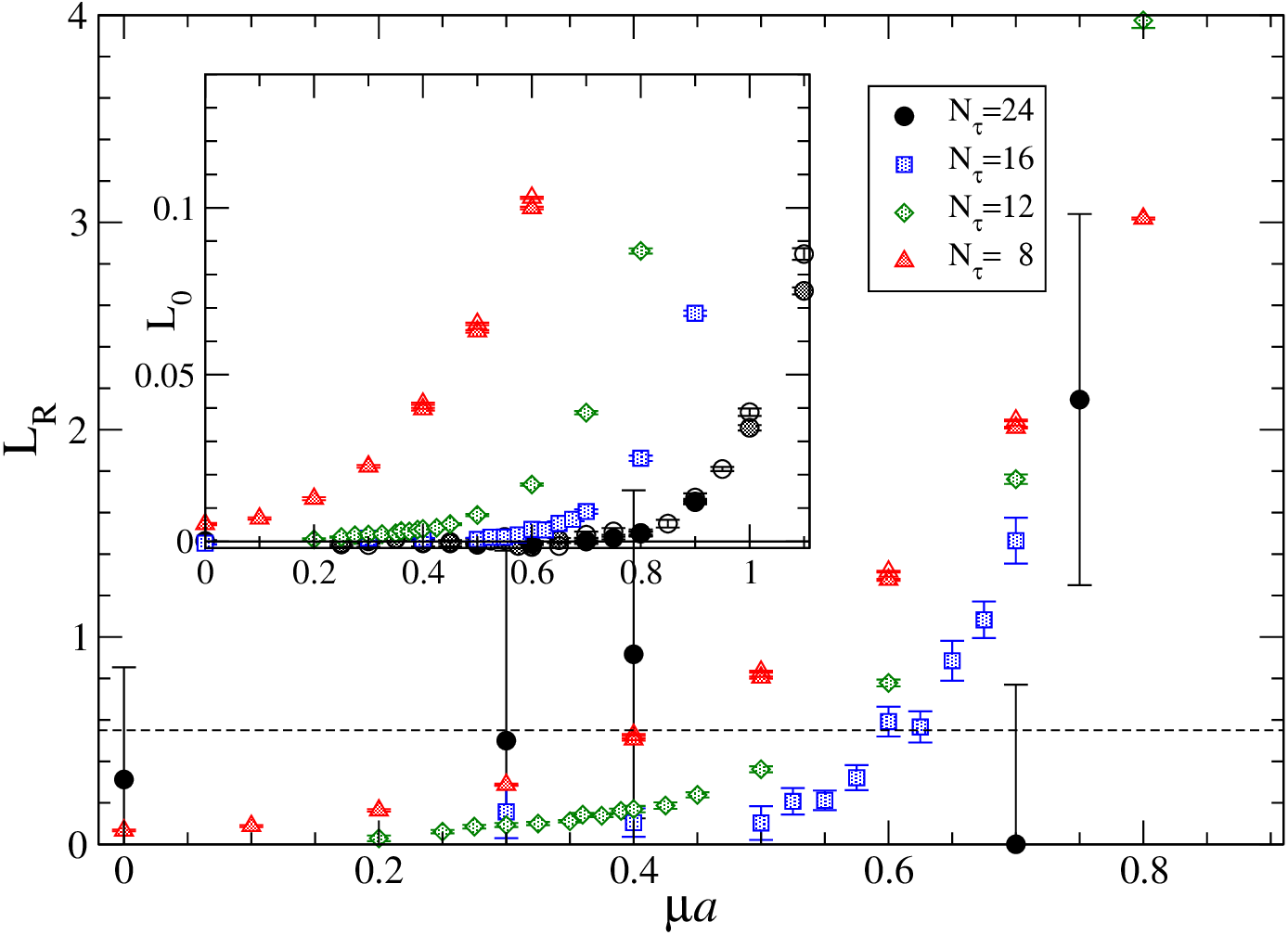}
\caption{Left: The diquark condensate $\bra qq\ket/\mu^2$
  extrapolated to $j=0$ for $N_\tau=24, 12, 8$ ($T=47,94,141$ MeV).
Right: The renormalised Polyakov loop as a function of chemical
  potential, for all temperatures.  The shaded symbols are for
  $ja=0.04$; the open symbols are $ja=0.02$.  The filled black circles
  are the results for the $16^3\times24$ lattice.  The dashed line
  indicates the inflection point of $L$ at $\mu=0$. The inset shows the
  unrenormalised Polyakov loop.}
\label{fig:orderparams}
\end{figure}

The left panel of fig.~\ref{fig:orderparams} shows the diquark condensate 
$\qq = \bra\psi^{2tr}C\gamma_5\tau_2\psi^1
  -\bar\psi^1C\gamma_5\tau_2\bar\psi^{2tr}\ket$
as a function of chemical potential, for the $N_\tau=24, 12$ and 8
lattices.  In the case of a weakly-coupled BCS condensate at the Fermi
surface, the diquark condensate, which is the number density of Cooper
pairs, should be proportional to the area of the Fermi surface, ie
$\qq\sim\mu^2$.

For the lowest temperature, $T=47$ MeV ($N_\tau=24$), we see that
$\qq/\mu^2$ has a plateau in the region $0.35\lesssim\mu a\lesssim0.6$.  The
increase for $\mu a\gtrsim0.6$ may be evidence of a transition to a
new state of matter at high density, although the impact of lattice
artefacts cannot be excluded.  The
lower limit of the plateau roughly coincides with the onset chemical
potential $\mu_o\approx m_\pi/2\approx0.33a^{-1}$, below which both
the quark number density and diquark condensate are expected to be
zero.  We find no substantial volume
dependence at any $\mu$.  Our results at $T=70$ MeV (not shown here)
are almost identical to those at $T=47$ MeV.
At $T=94$ MeV ($N_\tau=12$), $\qq$ is significantly suppressed, and
drops dramatically for $\mu a\gtrsim0.7$.  At $T=141$ MeV ($N_\tau=8$)
the diquark condensate is zero at all $\mu$, confirming that the
system is in the normal phase at this temperature.
%  A systematic
%investigation including more temperatures and an extrapolation to
%$j=0$ at all temperatures will be required to establish the exact
%location and nature of the superfluid to normal transition.

\begin{figure}[tb]
\includegraphics*[width=\colw]{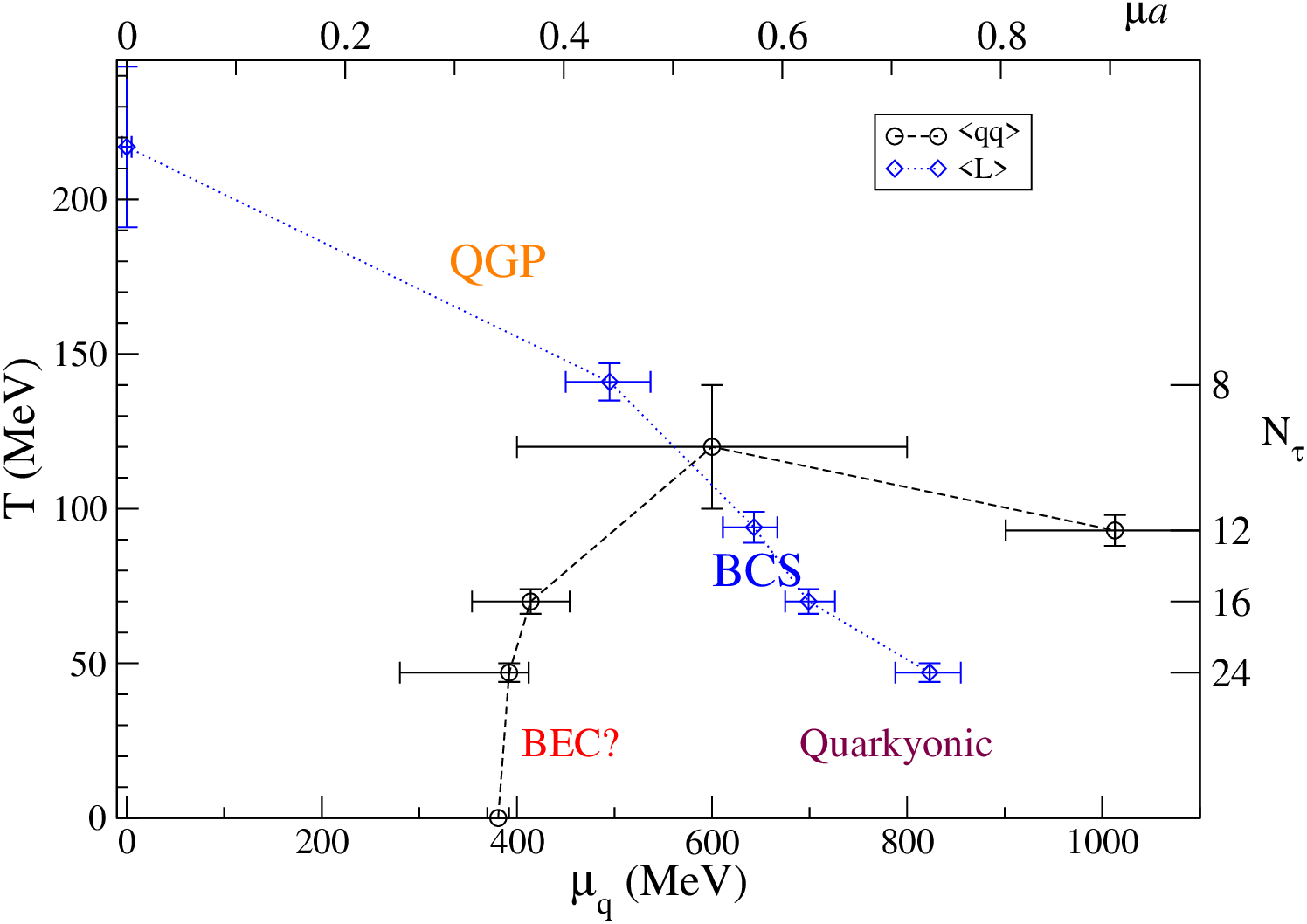}
\includegraphics*[width=\colw]{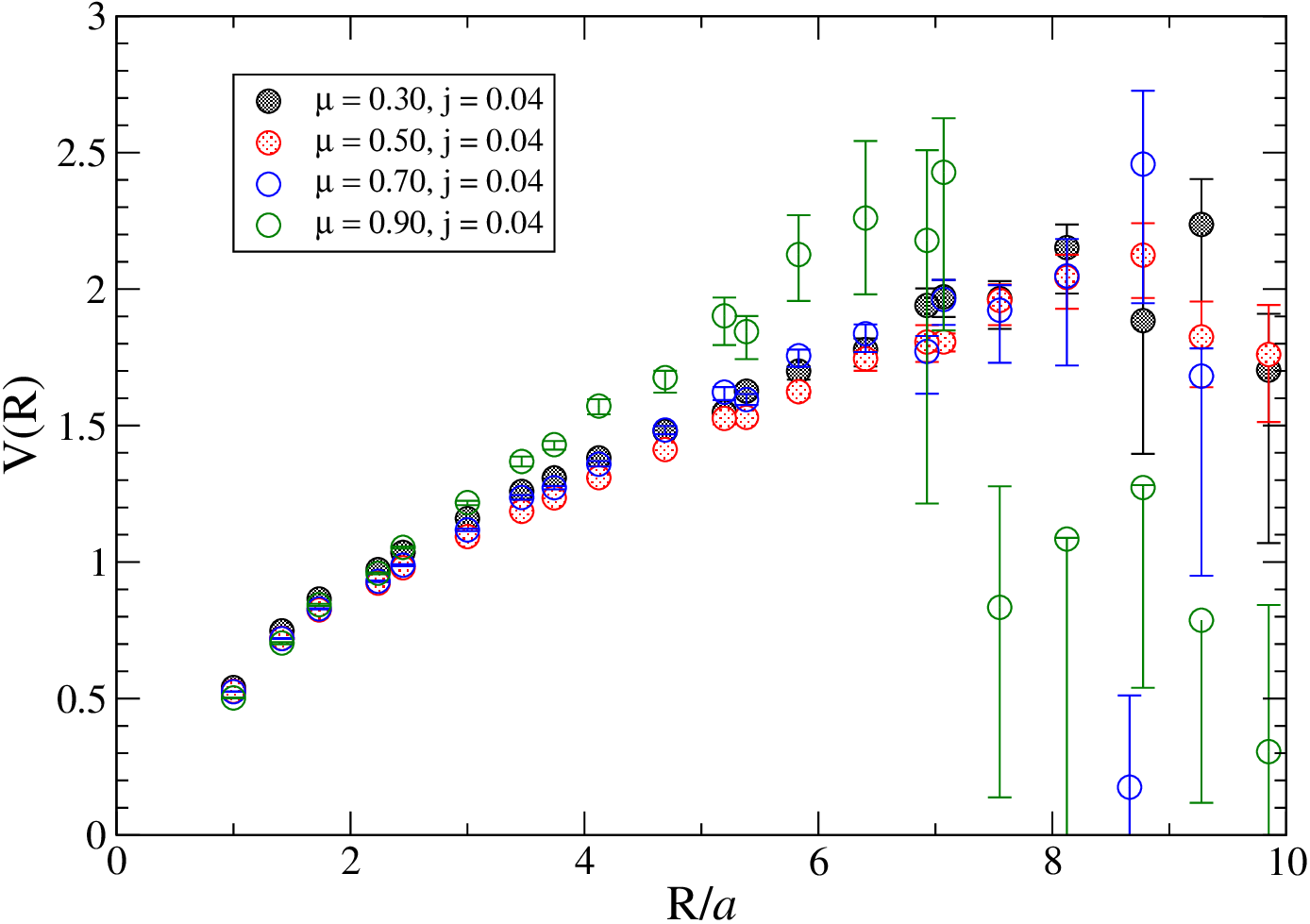}
\caption{Left: A tentative phase diagram, including the location of the
  deconfinement transition in the $(\mu,T)$ plane, determined from the
  renormalised Polyakov loop, and the transition to the diquark condensed $\bra
  qq\ket\neq0$ phase.  Right: The static quark potential computed from the
  Wilson loop, for the $12^3\times24$ lattice and different values of $\mu$.}
\label{fig:phases}
\end{figure}

In the right panel of fig.~\ref{fig:orderparams} we show the order
parameter for deconfinement, 
the Polyakov loop $L$, for our four temperatures. It
has been renormalised by requiring $L(Ta=\quarter,\mu=0)=1$, see
\cphases\ for details.
We see that for each temperature $T$, $L$ increases rapidly
from zero above a chemical potential $\mu_d(T)$ which we may identify
with the chemical potential for deconfinement.  In the absence of a
more rigorous criterion, we have taken the point where $L$ crosses the
value it takes at $T_d(\mu=0)$, $L_d=0.6$ \cphases, to define $\mu_d(T)$.  The
results are shown in fig.~\ref{fig:phases}, with error bars denoting
the range $L_d=$0.5--0.7.  To more accurately locate the deconfinement
line, we will need to perform a temperature scan for fixed
$\mu$-values. This is underway.

The estimates of critical chemical potentials for deconfinement and
superfluidity can be translated into a tentative phase diagram, shown in
fig.~\ref{fig:phases}.
In summary, from the order parameters we find signatures of three
different regions (or phases): a normal (hadronic) phase with
$\qq=0,\braket{L}\approx0$; a BCS (quarkyonic) region with
$\qq\sim\mu^2$ at low $T$ and intermediate to large $\mu$; and a
deconfined, normal phase with $\qq=0,\braket{L}\neq0$ at large $T$
and/or $\mu$.
After extrapolating our results to $j=0$ we see no
evidence of a BEC region described by $\chi$PT, with
$\qq\sim\sqrt{1-\mu_o^4/\mu^4}$ \cite{Kogut:2000ek}, in contrast with
earlier work with staggered lattice fermions \cite{Hands:2000ei}.
This may be related to the large value of $m_\pi/m_\rho$ in this
study.  Simulations with lighter quarks may yield further insight into this.

In the right panel of fig.~\ref{fig:phases} we show the static quark
potential computed from the Wilson loop at $N_\tau=24$, for $\mu a=0.3, 0.5, 0.7,
0.9$.  We find that as we enter the superfluid region, the string
tension is slightly reduced, but that this is reversed as $\mu$ is
increased further, leading to a strongly enhanced string tension at
$\mu a=0.9$, which according to our analysis of the Polyakov loops
should be in the deconfined region.  This agrees with the pattern that was already observed
in \cdeconf.  We also find no significance $j$-dependence in our
results.  At present we do not have a good understanding of why the
static quark potential should become antiscreened at large $\mu$.
Computing the static quark potential using Polyakov loop correlators
rather than Wilson loops may yield further insight into this issue.

\section{Equation of state}
\label{sec:eos}

\begin{figure}[tb]
\includegraphics*[width=\colw]{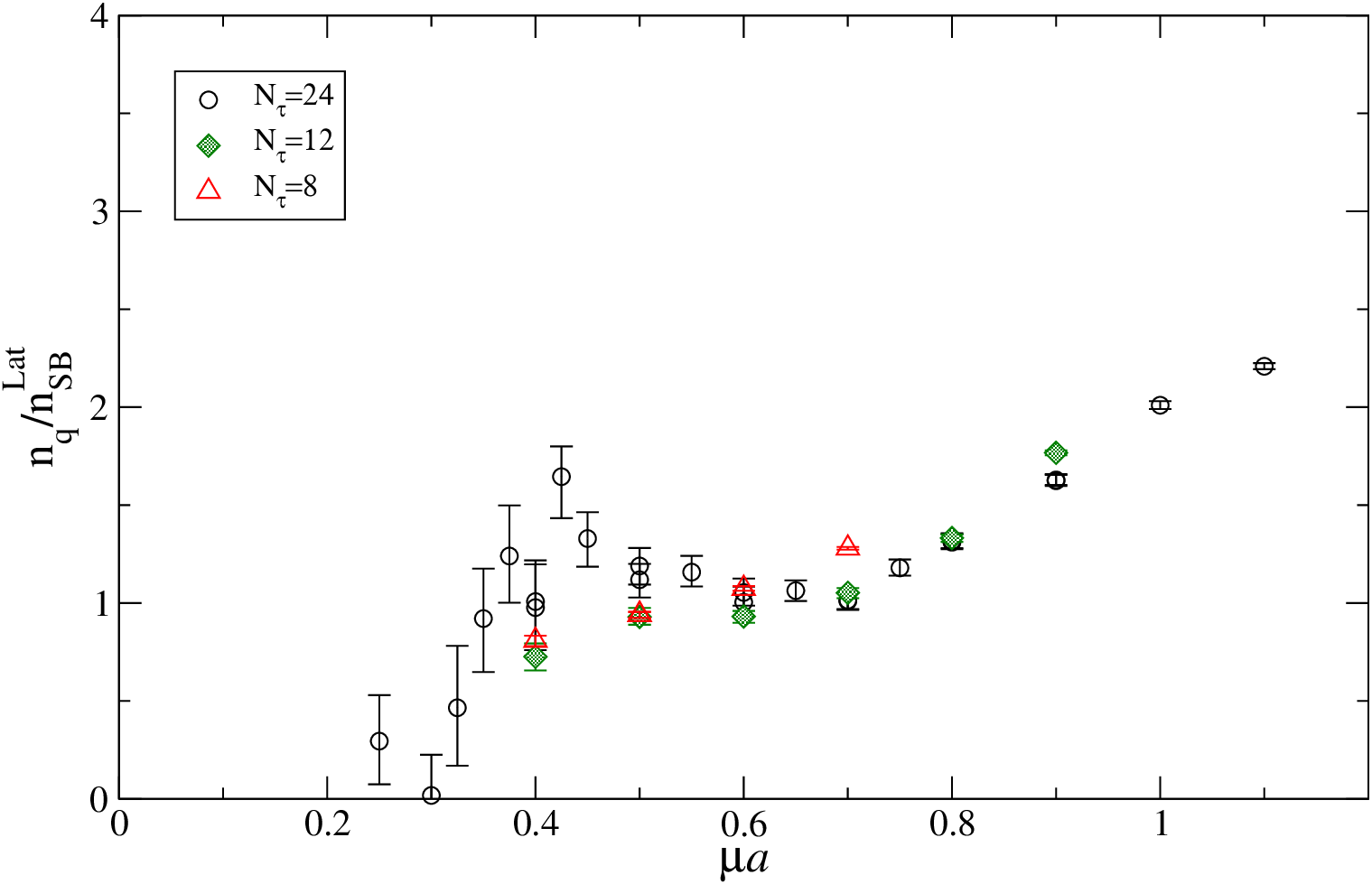}
\includegraphics*[width=\colw]{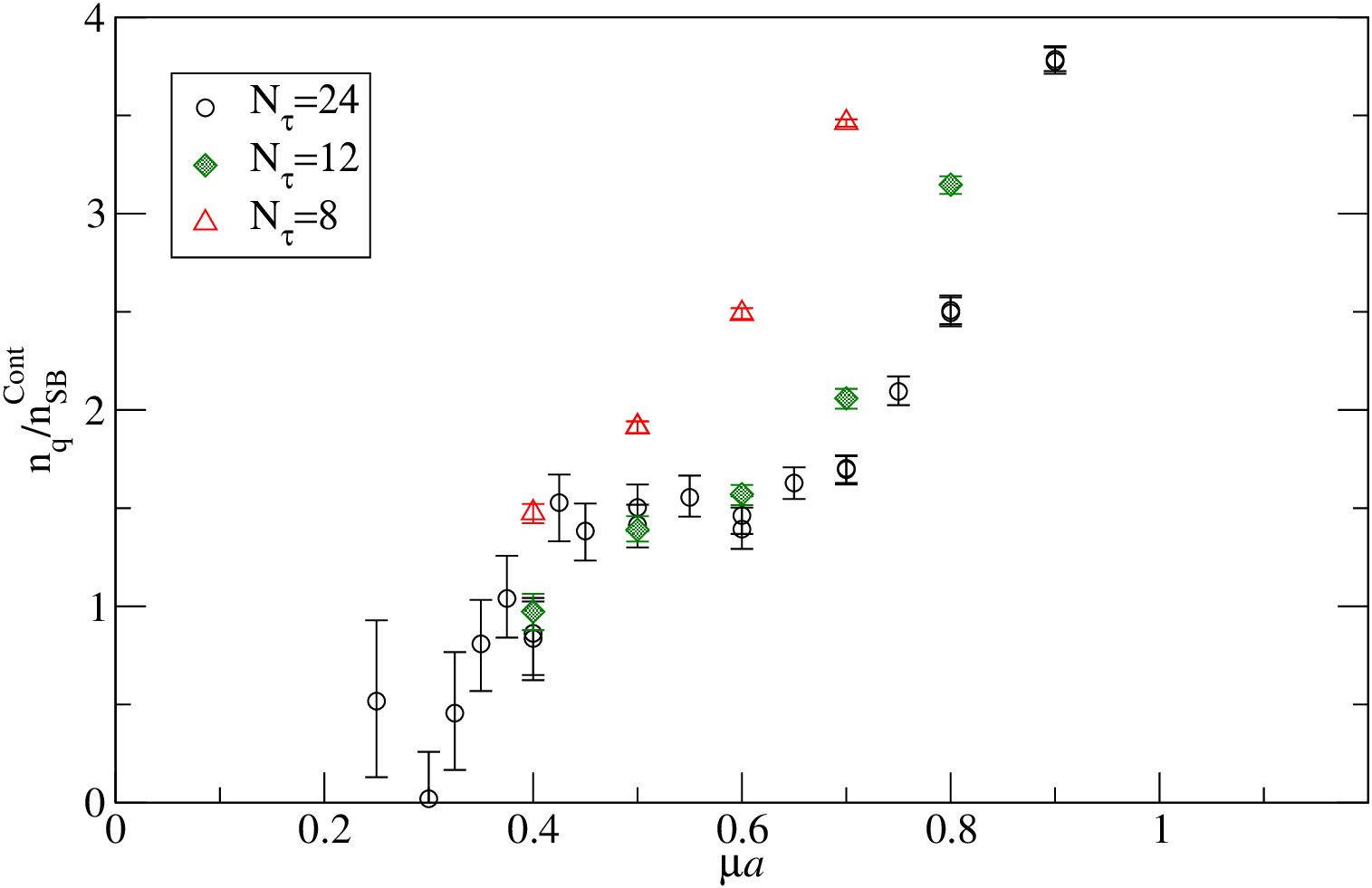}
\caption{The quark number density divided by the density for a
  noninteracting gas of lattice quarks (left) and continuum quarks
  (right).}
\label{fig:density}
\end{figure}

We now turn to the bulk thermodynamics of the system, and in
particular the quark number $n_q$ and the energy density $\eps$.
Fig.~\ref{fig:density} shows the quark number density $n_q$ for
$N_\tau=24, 12$ and 8, extrapolated to zero diquark source, and
normalised by the noninteracting value for lattice fermions on the
left and for continuum fermions on the right.  The difference between
the two gives an indication of the lattice artefacts.  We see that the
density rises from zero at $\mu\approx\mu_o=0.32a^{-1}$, and for the
two lower temperatures is roughly constant and approximately equal to
the  noninteracting fermion density in the region $0.4\lesssim\mu
a\lesssim0.7$. The peak  at $\mu a\simeq0.4$ in the $N_\tau=24$ data
in the upper panel is an artefact of the normalisation with $n_{SB}$
for a finite lattice volume: the raw numbers for the
$12^3\times24$ and $16^3\times24$ lattices are identical within
errors, but $n_{SB}$ differs by about 50\% around $\mu a=0.4$.
%We therefore conclude that our previous
%interpretation \cdeconf\ of the peak in $n_q/n_{SB}$ in this region as
%evidence of a BEC condensate described by $\chi$PT was probably
%erroneous.

The density for $N_\tau=8$ does not show any plateau as a function of
$\mu$; instead, $n_q/n_{SB}$ shows a roughly linear increase in the
region $0.4\leq\mu a\leq0.7$.  This is suggestive of the system being
in a different phase at this temperature.  We also note that
$n_q/n_{SB}$ for $N_\tau=12$ rises above the corresponding $N_\tau=24$
data for $\mu a\gtrsim0.7$, where, according to the results of
Sec.~\ref{sec:phases}, the hotter system is entering the deconfined,
normal phase.

These results lend further support to our previous conjecture that in
the intermediate-density, low-temperature region the system is in a
``quarkyonic'' phase: a confined phase (all excitations are
colourless) that can be described by quark degrees of freedom.

The renormalised energy density can be derived by going to an
anisotropic lattice formulation with bare anisotropies
$\gamma_g=\sqrt{\beta_t/\beta_s}, \gamma_q=\kappa_t/\kappa_s$ and
physical anisotropy $\xi=a_s/a_\tau$.
In the isotropic limit $\gamma_q=\gamma_g=\xi=1$ 
the energy density is then
given by $\eps=\eps_g+\eps_q$ with
\begin{align}
\eps_g &= \frac{3}{2a^4}\biggl[
\bra\Re\tr U_{ij}\ket
 \left(\del{\beta}{\xi}-\beta\del{\gamma_g}{\xi}\right)
 + \bra\Re\tr U_{i0}\ket
 \left(\del{\beta}{\xi}+\beta\del{\gamma_g}{\xi}\right)\biggr]\,,
\label{eq:epsG} \\
\eps_q &= \frac{1}{a^4}\bigg[
 \kappa^{-1}\del{\kappa}{\xi}\big(16+\pbp\big)
 - \kappa\del{\gamma_q}{\xi}\bra\psibar D_0\psi\ket\bigg]\,.
\label{eq:epsQ}
\end{align}
We have determined the Karsch coefficients $\dell{c_i}{\xi}$ with
$c_i=\gamma_g,\gamma_q,\beta,\kappa$ by performing simulations with
$\gamma_q, \gamma_g\neq1$.  Our estimates for these coefficients are
\cphases\
\begin{equation}
\del{\gamma_g}{\xi} = 0.90\err{4}{14},\quad
\del{\gamma_q}{\xi} = 0.13\err{40}{5},\quad
\del{\beta}{\xi} = 0.59\err{0.24}{1.37},\quad
\del{\kappa}{\xi} = -0.052\err{69}{15}\,.
\end{equation}
\begin{figure}[tb]
\includegraphics*[width=\textwidth]{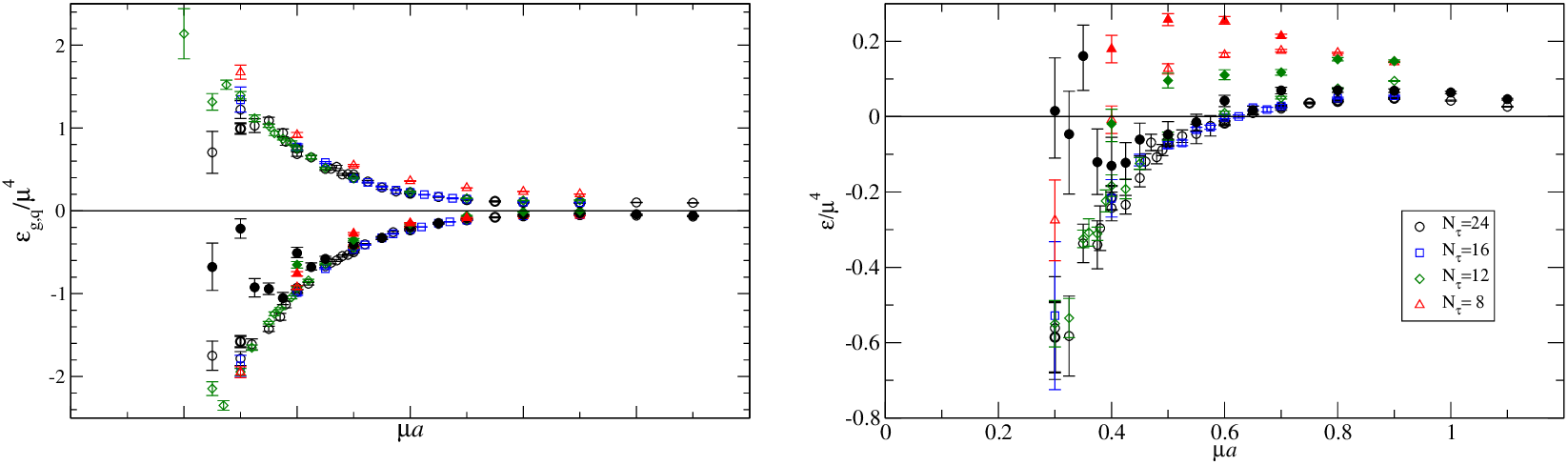}
\caption{The quark and gluon contributions to the energy density
  (left) and total energy density (right), divided by $\mu^4$, for
  $ja=0.04$ (open symbols) and $j=0$ (filled symbols).}
\label{fig:energy}
\end{figure}

Our results for the energy density are shown in
fig.~\ref{fig:energy}.  We see that the quark contribution is negative
for all values of $\mu$ and $T$, but this is balanced by the positive
gluon contribution, giving a positive or zero total energy.  
The energy density is very sensitive to the values of the Karsch
coefficients \cphases; for example, if $\dell{\gamma_q}{\xi}$ is
changed from the suprisingly low value of 0.13 to a more `natural'
value of 0.8, we find that $\eq>0$ for $\mu a\gtrsim 0.6$.

\section{Heavy quarkonium}
\label{sec:heavy}

\begin{figure}
\begin{center}
\includegraphics*[width=\colw]{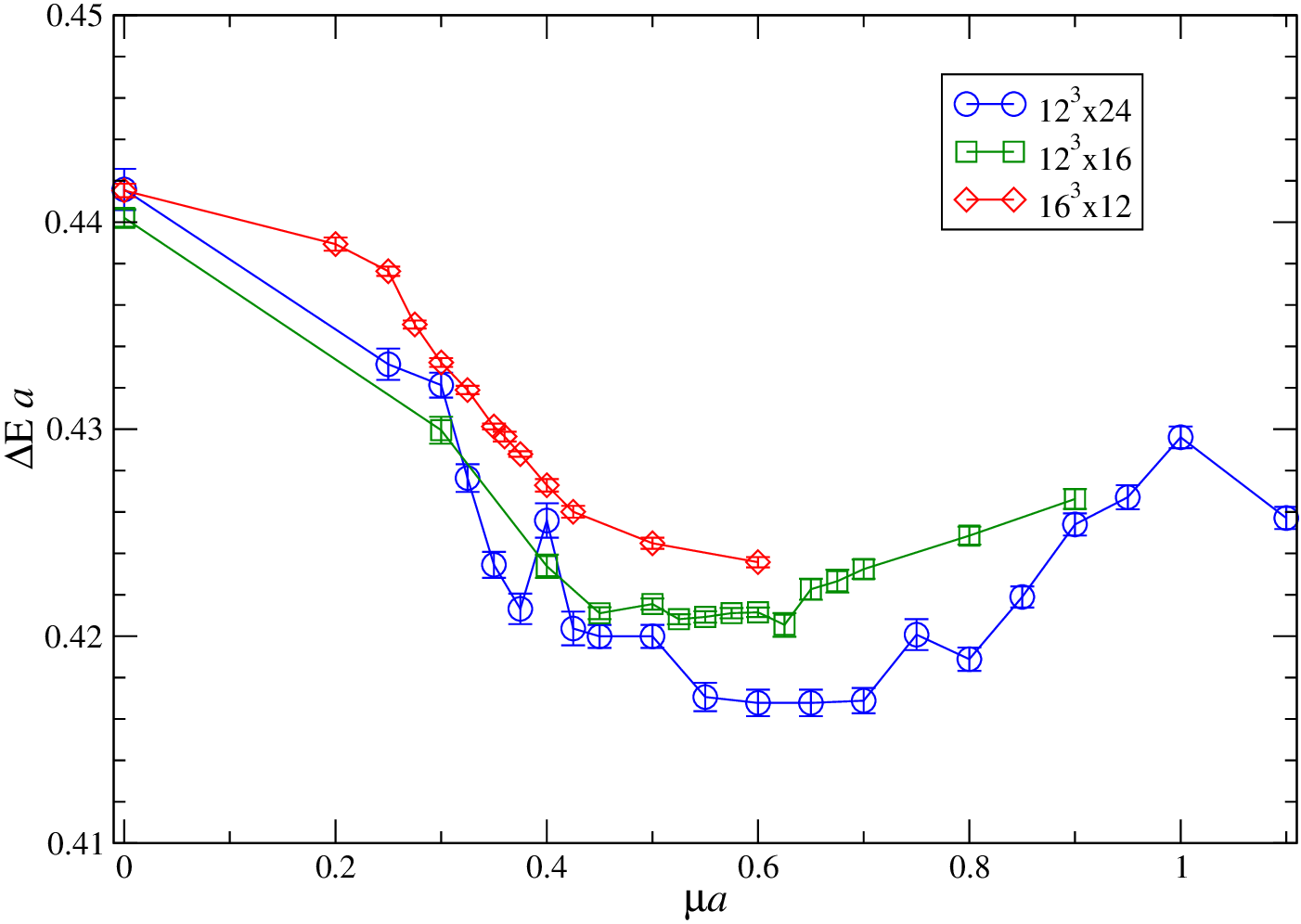}
\includegraphics*[width=\colw]{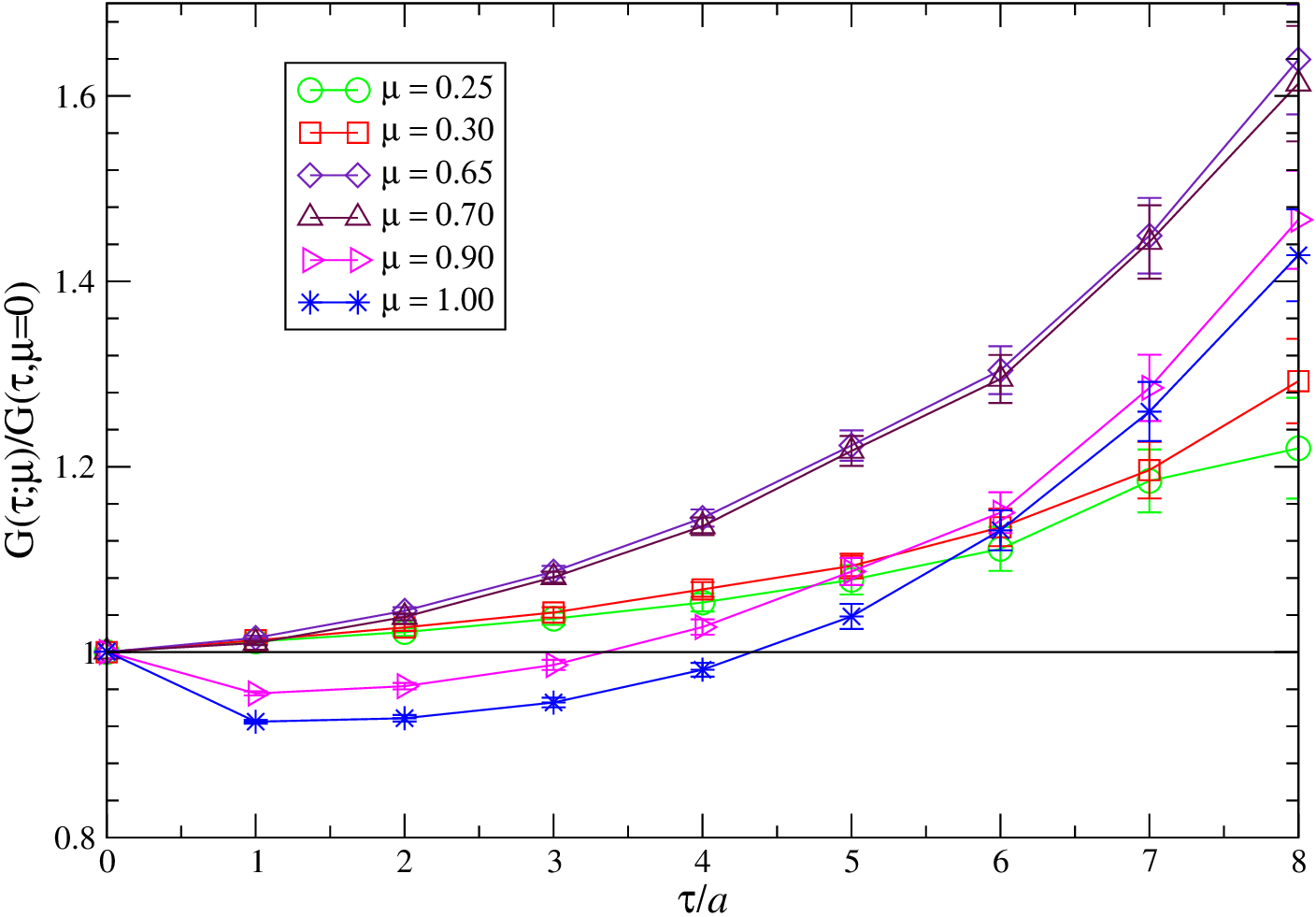}
\end{center}
\caption{Left: Temperature dependence of the $^1S_0$ state energy
  vs. $\mu$ for $Ma = 5.0$ with $j = 0.04$.  Right: The ratio
  $\sum_\xv G(\xv,\tau;\mu)/\sum_\xv G(\xv,\tau;0)$ for $^1P_0$
  correlators on $12^3 \times 24$ with $Ma = 5.0$. Due to the
  noisiness of the P-wave data, only a limited $\tau$ range is shown.}
\label{fig:quarkonium} 
\end{figure}

We have investigated the heavy quarkonium spectrum by computing
non-relativistic QC$_2$D correlators on our $N_\tau=24$, 16 and 12
lattices.  We use an $\order(v^4)$ lattice NRQCD lagrangian
\cite{Bodwin:1994jh} to compute the heavy quark Green function; see
\cite{Hands:2012yy} for further details.  We find that the S-wave
correlators can be fitted with an exponential
decay $\propto e^{-\Delta E_n\tau}$
even once $\mu\neq0$; moreover the fits are quite stable over large
ranges of $\tau$, indicating that $S$-wave bound states persist
throughout the region 47 MeV $\lesssim T\lesssim$ 90 MeV.

Fig.~\ref{fig:quarkonium} shows the $T$- and $\mu$-dependences of the
$^1S_0$ state energy $\Delta E$. 
We see that as $\mu$ is varied, initially the $^1S_0$ state energy
decreases from that at $\mu = 0$, but once $\mu$ reaches the region
$\mu_1 (\simeq 0.5 )\le \mu a \le \mu_2 (\simeq 0.85)$, the $^1S_0$
state energy stays roughly constant. For $\mu > \mu_2$, the $^1S_0$
state energy starts increasing again.  
In contrast to the observables studied in Secs~\ref{sec:phases} and
\ref{sec:eos}, we find no clear, systematic dependence on the diquark
source term for $\mu a\lesssim0.5$.  For $\mu a\gtrsim0.5$ on the
other hand, $\Delta E(ja=0.02)<\Delta E(ja=0.04)$.  This suggests that
the energy, extrapolated to $j=0$, may continue to decrease up to $\mu
a\approx0.7$ before increasing.

As the temperature increases from 47 MeV ($N_\tau=24$) to 70 MeV
($N_\tau=16$) we find that the point where the energy of the $^1S_0$
state starts increasing goes from $\mu a\approx0.7$ to 0.55.  This is
consistent with the estimate of the deconfinement transition in
Sec.~\ref{sec:phases}.  For $N_\tau=12$ we do not yet have any data in
the $\mu$-region which might confirm this.  It is interesting to note
that $\Delta E$ increases with increasing $T$, in accordance with
what has been observed in hot QCD with
$\mu=0$~\cite{Aarts:2011sm}.

In contrast to the $S$-waves, it is difficult to find stable
exponential fits to the $P$-wave correlators with the current
Monte-Carlo data before statistical noise sets in, except for the case
$\mu a \le 0.25$. In the right panel of fig.~\ref{fig:quarkonium} we
instead show the ratios of the
$^1P_0$ correlators at different values $\mu\not=0$
to the correlator at $\mu=0$.  Note that any effect we observe
is entirely due to the dense medium.

The S-wave correlator ratios show an increase with $\tau$ which
corresponds to the
negative $^1S_0$ energy difference $\Delta E(\mu) - \Delta E(\mu-0)$
that was previously observed.  In the quarkyonic region, the
$P$-wave ratios behave similarly to the $S$-wave, but in the
deconfined region ($\mu \ge \mu_2$), the $P$-wave ratios
are non-monotonic, initially decreasing with $\tau$ before turning to
rise above unity for $\tau/a\sim4$. On the other hand, the 
P-wave correlator ratios on the $12^3 \times 16$ and $16^3 \times 12$
lattice show monotonic behavior similar to those of the S-waves,
suggesting a subtle interplay of density and temperature effects
on the P-wave states.

\section{Summary and outlook}

From lattice simulations of dense QC$_2$D at a range of temperatures,
we have identified three distinct regions of the phase diagram: a
hadron gas at low $\mu$ and $T$, a quarkyonic region at intermediate
$\mu$ and low to intermediate $T$, and a deconfined quark--gluon
plasma at high $T$ and/or $\mu$.  Taking the limit of zero diquark
source has served to make our identification of the quarkyonic region
more robust.  Investigations into the exact nature and location of the
deconfinement and the superfluid to normal transitions are underway,
as are simulations at smaller lattice spacings and with smaller quark
masses.

\section*{Acknowledgments}

This work is carried out as part of the UKQCD collaboration and the
DiRAC Facility jointly funded by STFC, the Large Facilities Capital
Fund of BIS and Swansea University.  We thank the DEISA Consortium
(www.deisa.eu), funded through the EU FP7 project RI-222919, for
support within the DEISA Extreme Computing Initiative, and the USQCD
for use of computing resources at Fermilab.  JIS and SC
acknowledge the support of Science Foundation Ireland grants
08-RFP-PHY1462, 11-RFP.1-PHY3193 and 11-RFP.1-PHY3193-STTF-1.
SK is grateful to STFC
for a Visiting Researcher Grant and is supported by the National Research
Foundation of Korea grant funded by the Korea government (MEST) No.\
2011-0026688. DM acknowledges support from the U.S. Department of
Energy under contract no. DE-FG02-85ER40237. 
JIS acknowledges the support and hospitality of the Institute for
Nuclear Theory at the University of Washington, where part of this
work was carried out. 

\bibliography{density,lattice,hot}
\end{document}